\documentclass[aps,twocolumn,showpacs]{revtex4}
\usepackage{graphicx,epsfig}
\renewcommand{\vec}[1]{{\mathbf #1}}
\newcommand{\calM}{{\mathcal M}}
\newcommand{\epstil}{\tilde{\epsilon}}
\newcommand{\omtil}{\tilde{\omega}}
\begin{document}

\title{Polariton condensation with localised excitons and propagating photons}
\author{Jonathan Keeling, P.~R.~Eastham, M.~H.~Szymanska
and P.~B.~Littlewood}

\affiliation{Cavendish Laboratory, Madingley Road, Cambridge CB3 OHE, U.K.}
\pacs{71.36.+c,42.55.Sa,03.75.Kk}

\begin{abstract}
  We estimate the condensation temperature for microcavity polaritons,
  allowing for their internal structure. We consider polaritons formed
  from localised excitons in a planar microcavity, using a generalised
  Dicke model. At low densities, we find a condensation temperature
  $T_c \propto \rho$, as expected for a gas of structureless
  polaritons. However, as $T_c$ becomes of the order of the Rabi
  splitting, the structure of the polaritons becomes relevant, and the
  condensation temperature is that of a B.C.S.-like mean field theory.
  We also calculate the excitation spectrum, which is related to
  observable quantities such as the luminescence and
  absorption spectra.
\end{abstract}
\maketitle


Bose condensation of polaritons in two-dimensional semiconductor
microcavities has been the object of continuing experiments.
Recent progress includes observations
of a nonlinear increase in the occupation of the ground state
\cite{dang98:_stimul_polar_photol_semic_microc,yamamoto02:_condensation},
sub-thermal second order coherence
\cite{yamamoto02:_condensation}, changes to the momentum distribution
of polaritons \cite{deng03:_polar}, and stimulated processes in
resonantly pumped
cavities\cite{savvidis00:_angle_reson_stimul_polar_amplif}.
However, unambiguous evidence for the existence of a polariton
condensate is still lacking. Theoretical predictions of the conditions
required for condensation, and of signatures of condensation, are
therefore of great interest.
We find that the momentum distribution of radiation from the cavity
shows a dramatic change when condensed , which, together with the gap
in the luminescence could provide such signatures.

Due to the small polariton mass, a weakly interacting bose gas of
polaritons predicts extremely high transition temperatures, which
provides the motivation for studying polariton condensation; see
\cite{kavokin03:_polar} and references therein.
At such high temperatures, when $T_c$ is of the order of
the Rabi splitting, the internal structure of the polariton becomes
relevant, and treating polaritons as structureless bosons fails.  This
can occur at density scales much less than the exciton saturation
(Mott) density; while current experiments \cite{deng03:_polar} are
well below the Mott density, they are at densities where the
internal structure is relevant.

In this paper, we study polariton condensation in thermal equilibrium,
including polariton internal structure.
%
Although current experiments may not have reached equilibrium, this
remains their aim; further the signatures of equilibrium
condensation are instructive when considering the non-equilibrium
case.
We consider a model of 
localised excitons in a two-dimensional microcavity, building on
previous studies of zero-dimensional
microcavities\cite{eastham01:_bose}. This is motivated by experimental
systems such as organic semiconductors \cite{lidzey99:_room}, quantum
dots\cite{woggon03:_dotindot}, and disordered quantum
wells\cite{hess94}.  Moreover it has been argued
\cite{szymanska03:_polar} that exciton localisation might aid
condensation; inhomogeneous broadening due to
localisation suppresses condensation less than
the homogeneous dephasing for propagating excitons.
Finally, since the exciton mass is much larger
than the photon mass, we expect much of the physics of condensation in
the localised exciton model to carry over to systems with propagating
excitons, for densities much less than the Mott density.


%
%

For simplicity, we model the localised excitons as isolated two-level
systems, with the two states corresponding to the presence or absence
of an exciton on a particular site. 
A two-level system will be represented as a Fermion occupying one of
two levels, further constrained so that the total occupation of a site
is one.
This constraint describes a hard core repulsion between excitons on a
single site; we neglect Coulomb interactions between different sites as
being small compared to the photon mediated interaction.
In a two dimensional microcavity, the photon dispersion relation for
low $k$ is $\hbar \omega_k=\hbar \omega_0 + \hbar^2 k^2/2m$, where $m=
(\hbar/c) (2\pi/w)$ for a cavity width $w$.
Each mode couples to all excitons, with different phases.
Describing the coupling in the electric dipole gauge, the coupling
strength is $g=d_{ab}\sqrt{2\pi \hbar \omega_k}$, in which we
approximate $\omega_k \approx \omega_0$, and $d_{ab}$ is the dipole
moment. 
These considerations lead to the generalised Dicke
Hamiltonian\cite{dicke54:_coher_spont_radiat_proces}
\begin{eqnarray}
  \label{eq:1}
    H &=&
    \sum_{j=1}^{j=nA} \epsilon_j \left
      (b^{\dagger}_j b^{}_j -a^{\dagger}_j a^{}_j 
    \right)
    +
    \sum_{k=2\pi l/\sqrt{A}} \hbar \omega_k \psi^{\dagger}_k \psi^{}_k
    \nonumber\\
    &+&
    \frac{g}{\sqrt{A}} \sum_{j,k} \left(
      e^{2\pi i \vec k \cdot \vec r_j} \psi_k a_j^{\dagger} b_j +
      e^{-2\pi i \vec k \cdot \vec r_j} \psi_k^{\dagger} b_j^{\dagger} a_j
    \right).
\end{eqnarray}
Here $A\rightarrow\infty$ is the quantisation area and $n$ the areal
density of two level systems, i.e. sites where an exciton may exist.
Without inhomogeneous broadening, the energy of a bound exciton is
$2\epsilon=\Delta+\omega_0$, defining the detuning $\Delta$ between
the exciton and the photon.
The grand canonical ensemble, $\tilde{H}=H-\mu N$, allows the
calculation of equilibrium for a fixed total number of excitations
(excitons and photons);
\begin{equation}
  \label{eq:2}
      N=  \sum_{j=1}^{j=nA} \frac{1}{2}\left(
      b^{\dagger}_j b^{}_j -a^{\dagger}_j a^{}_j  + 1
    \right)
    + \sum_{k=2\pi l/\sqrt{A}} \psi^{\dagger}_k \psi^{}_k.
\end{equation}
We therefore define $\omtil_k=\omega_k-\mu$ and
$\epstil=\epsilon-\mu/2$.
If $\mu=0$, this Hamiltonian is an extension of that studied by Hepp
and Lieb.\cite{hepp73:_equil_statis_mechan_matter_quant}, as discussed
further in \cite{eastham01:_bose}.

\begin{figure}[htbp]
  \centering
  \includegraphics[width=3.5in]{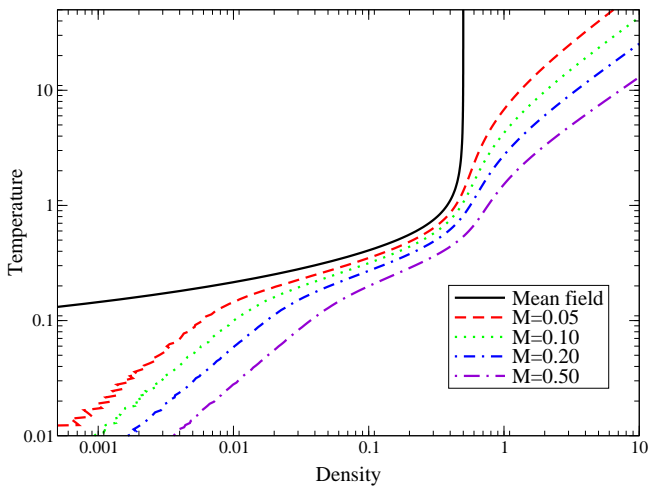}
  \caption{%
    Phase boundaries calculated from mean field theory, and including
    fluctuation corrections for four different values of photon mass.
    Temperature and density plotted in units of $g\sqrt{n}$ and $n$.
  }
  \label{fig:phasediag}
\end{figure}

Using the characteristic density, $n$ and energy scale $g\sqrt{n}$
(the Rabi splitting), the system is controlled by two dimensionless
parameters: Detuning, $\Delta^{\ast}=\Delta / g\sqrt{n}$ and Photon
mass, $m^{\ast}=m g / \hbar^2 \sqrt{n}$.
For the
system in ref.~\cite{yamamoto02:_condensation},  $g\sqrt{n}
\approx 20 \mathrm{meV}$, $\hbar \omega_0\approx 1.6 \mathrm{eV}$, and
$m\approx 10^{-5} m_{e}$.
For delocalised but interacting excitons, assuming
that the exciton-exciton interaction is a hard-core repulsion on the
scale of the Bohr radius gives an upper bound on the density of
available exciton sites as $n \approx 10^{11} n_{\mathrm{QW}} \mathrm{cm}^{-2}$, where $n_{\mathrm{QW}}$ is the number of quantum
wells in the cavity. These values lead to dimensionless parameters
$m^{\ast} \approx 10^{-3}$, and by design $\Delta^{\ast}=0$. The
quoted exciton densities in Ref. \cite{deng03:_polar} are $\approx
10^{9} \mathrm{cm}^{-2}$ per well, leading us to estimate the fraction
of occupied exciton sites to be less than $10^{-2}$.  Using these values,
our work suggests (see Fig.~\ref{fig:phasediag}) that current 
experiments~\cite{deng03:_polar} are in the interaction dominated
mean field regime.
%
%

Using the standard functional path integral techniques for the
partition function as in \cite{eastham01:_bose} and integrating over
the Fermion fields yields an effective action for photons:
\begin{eqnarray}
  \label{eq:3}
  S[\psi]&=&
  \int_0^{\beta} d\tau
  \sum_k
  \psi^{\dagger}_k \left(\partial_{\tau} + \hbar \omtil_k\right)  \psi^{}_k
  - 
  N \mathrm{Tr} \ln \left(\calM\right)
  \\
  \calM^{-1} &=&  
  \left(
    \begin{array}{ll}
      \partial_{\tau} + \epstil & 
      \frac{g}{\sqrt{A}}\sum_k e^{2\pi i\mathbf{k}\mathbf{r_n}} \psi_k \\
      \frac{g}{\sqrt{A}}\sum_k e^{-2\pi i\mathbf{k}\mathbf{r_n}} \bar{\psi_k}&
      \partial_{\tau} - \epstil 
    \end{array}
  \right)
  \nonumber.
\end{eqnarray}
We then proceed by minimising $S[\psi]$ with a static uniform $\psi$
and expanding around this minimum.
The minimum, $\psi_0$, satisfies the equation:
\begin{equation}
  \label{eq:4}
  \omtil_0 \psi_0
  = g^2 n
  \frac{\tanh ( \beta E )}{2 E} \psi_0,
  \quad
  E=\sqrt{\epstil^2 + g^2 n |\psi_0|^2}.
\end{equation}
which describe the mean-field condensate of coupled coherent photons
and exciton polarisation studied in \cite{eastham01:_bose}.

To consider the influence of incoherent fluctuations we 
expand the logarithm in eq.~(\ref{eq:3}) for fluctuations
$\psi=\psi_0 + \delta \psi$ and obtain a contribution to the action:
\begin{math}
  \frac{1}{2} \mathrm{Tr}(\calM \delta\calM^{-1} \calM \delta\calM^{-1}).
\end{math}
This, with the bare photon action, yields the inverse Greens function
for photon fluctuations ${\mathcal G}^{-1}(\omega,k)$.
The resultant expression has terms proportional to $\delta_{\omega}$.
These arise from 2nd order poles in the Matsubara summation;
such terms vanish on analytic continuation, and so do not affect
the retarded Greens function.

\begin{figure}[htbp]
  \centering
  \includegraphics[width=3.5in]{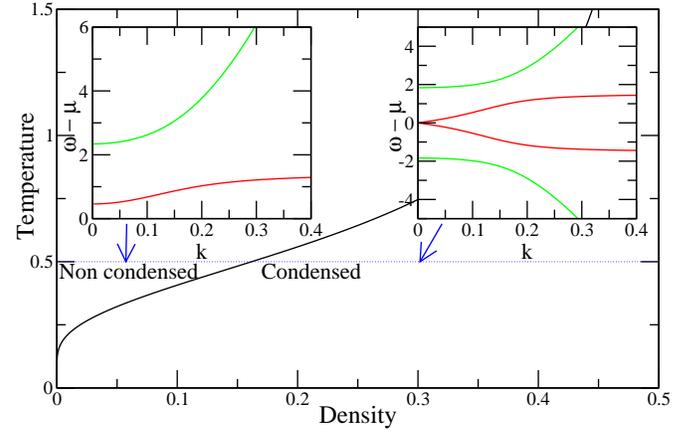}
  \caption{Condensed and uncondensed spectra, superimposed on the
    mean field phase diagram to show choice of mean field density
    and temperature.
    (Parameters $\Delta^{\ast}=0$, $m^{\ast}=0.01$. $\mu/g\sqrt{n}=-1.4$ below
    transition, $\mu/g\sqrt{n}=-0.6$ above.  Energies and densities plotted
    in units of $g\sqrt{n}$ and $n$.)
  }
  \label{fig:spectrum}
\end{figure}

The Greens function has poles, representing the energy of excitations
w.r.t.\ the chemical potential.
In the normal state there are two poles which describe the upper and
lower polariton dispersion
\begin{equation}
  \label{eq:5}
    E_{\pm}=
  \frac{1}{2}
  \left(
    (\omtil_k + 2 \epstil)
    \pm
    \sqrt{%
      (\omtil_k - 2 \epstil)^2
      + 4 g^2 n \tanh(\beta\epstil)
    }
  \right).
\end{equation}
and are presented in  Figure~\ref{fig:spectrum}.
The polariton dispersion~(\ref{eq:5}) from localised excitons has the
same structure as from propagating excitons, since the photon
dispersion dominates.
In the condensed state the excitation spectrum changes dramatically
(see Fig~\ref{fig:spectrum}).
There are four poles, reflecting the symmetry of adding or subtracting
a particle in the presence of the condensate:
\begin{eqnarray}
  \label{eq:6}
  \xi_{1,2}^2 &=& \frac{1}{2} \left(
    A(k) \pm 
    \sqrt{A(k)^2 - 
      16\frac{k^2}{2m} \left(
        \frac{E^2 k^2}{2m} + B
      \right)
    } \right)
  \nonumber\\
  A(k)&=&4E^2 + \omtil_k^2 + 4\epstil\omtil_0
  ,\quad
  B=g^2 n \frac{\left|\psi\right|^2}{N}\omtil_0.
\end{eqnarray}
At low $k$, the inner lines correspond to changing the phase of the
condensate.
At higher $k$, unlike in B.C.S. superconductivity, amplitude and phase
modes get mixed.

For small momenta, the phase mode is linear, $\xi_1=\pm c k + O(k^2)$.
The phase velocity has the form
\begin{equation}
  \label{eq:7}
  c= \sqrt{%
    \frac{1}{2m}
    \left(\frac{4\omtil_0 g^2 n}{\xi_2(0)^2}\right)
    \left(\frac{|\psi_0|^2}{N}\right)}
  \approx \sqrt{\frac{\lambda}{2m}\frac{\rho_0}{n}}
\end{equation}
where the second expression is for comparison the form for a dilute
Bose gas with contact interaction $\lambda$.
The phase velocity increases for small $\psi_0$ then decreases as
saturation of the excitons reduces the effective exciton-photon
interaction strength.

\begin{figure}[htbp]
  \centering
  \includegraphics[width=3.5in]{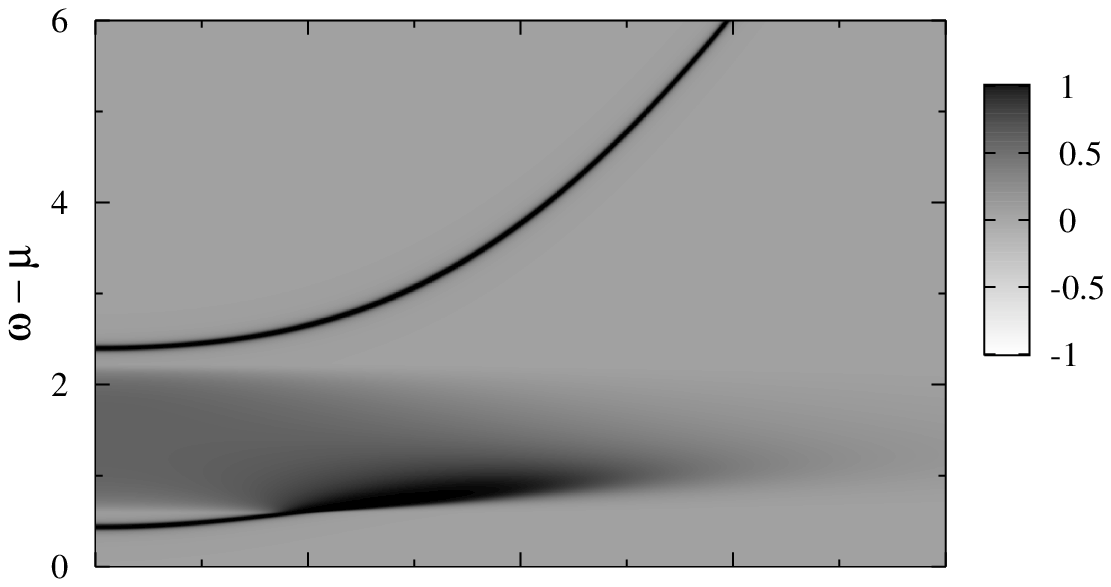}
  \includegraphics[width=3.5in]{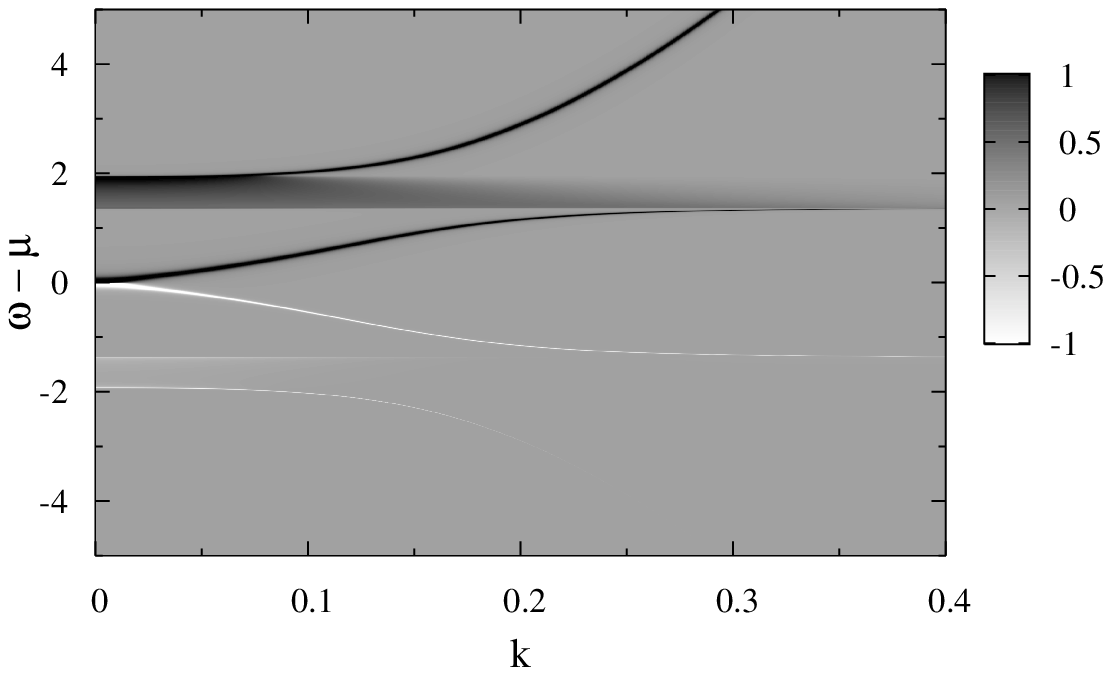}
  \caption{%
    Absorption coefficient v.s.\ momentum (x axis) and energy (y axis)
    calculated above (top) and below (bottom) phase transition.
    Negative values indicate gain.
    (Parameters as for figure~\ref{fig:spectrum}, inhomogeneous
    broadening width $0.3g\sqrt{n}$.)
    The colour scale covers the range $-\infty$ to $\infty$.
  }
  \label{fig:dos}
\end{figure}

The difference in the structure of the excitation spectra between
condensed and uncondensed states may be seen in the luminescence
spectra and in the angle-resolved luminescence $N(k)$.
Although four poles exist, they may have very different spectral
weights, and would therefore show up differently in optical
measurement.
This can be understood more clearly after introducing inhomogeneous
broadening of the exciton energies, which broadens the poles.
In this case, one may plot the spectral weight of the photon greens
function, $2 \Im\left[\mathcal{G}(i\omega=z+i\eta)\right]$ as
$\eta\rightarrow 0$, which corresponds experimentally to the
absorption coefficient.
Figure~\ref{fig:dos} shows this for the same parameters as
figure~\ref{fig:spectrum}.

In the condensed case, there are three bright lines above the chemical
potential; the Goldstone mode (which is not broadened as it lies
inside the gap for single particle excitations); the edge of the gap,
and the broadened upper polariton mode.
If the inhomogeneous broadening is large, the gap edge peak 
and upper polariton will be broadened.
This will result in a band of incoherent luminescence separated from
the coherent emission at the chemical potential by the gap.
If the broadening is small w.r.t.\ the gap then very little structure
will be seen.
Such structure can however still be seen at low densities.

\begin{figure}[htbp]
  \centering
  \includegraphics[width=3.5in]{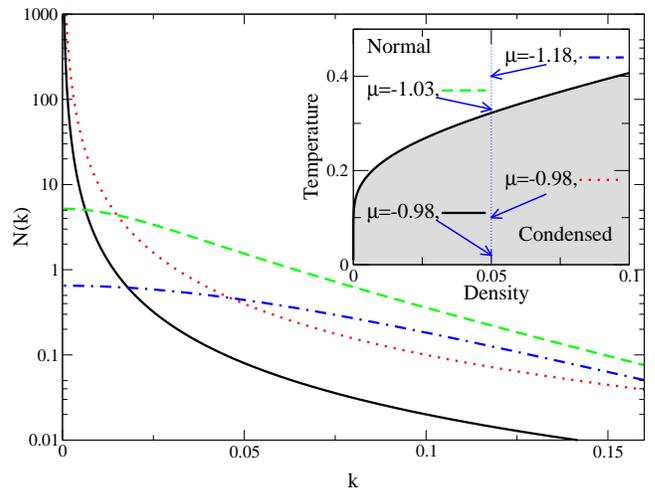}
  \caption{Momentum distribution of photons, from which follows the 
    angular distribution.  Two uncondensed, and two  condensed distribution
    are shown, for mean field densities as shown in the inset.
    (Parameters $\Delta^{\ast}=0$, $m^{\ast}=0.01$.)}
  \label{fig:nofk}
\end{figure}

The momentum distribution of photons may also be found from the photon
Greens function.
This is characteristic of the angular distribution of radiation; if
reflectivity of the mirrors were independent of angle, and
approximating $\omega_k\approx\omega_0$ then $\sin(\theta)=c
k/\omega_0$.
The momentum distribution is given by:
\begin{eqnarray}
  \label{eq:8}
  N(k) &=& 
  \lim_{\eta\rightarrow0}
  \left< \psi^{\dagger}_k(\tau + \eta) \psi^{}_k(\tau) \right> 
\end{eqnarray}
Although the system is two dimensional, a condensed momentum
distribution can be calculated at low temperatures by considering only
phase fluctuations~\cite{keeling04:_angul}.
Figure~\ref{fig:nofk} shows this for both condensed and uncondensed
states.
The linear phase mode leads to a power law divergence at low momenta, becoming
sharper as the temperature decreases.

We now estimate the phase boundary including the effects of
fluctuations.
%
%
We calculate the free energy from the action in eq.~(\ref{eq:3}),
evaluated to Gaussian order.
Taking the total derivative of free energy w.r.t.\ chemical potential
gives the total density, and evaluating this at the critical chemical
potential gives an estimate of the phase boundary.
This follows the method of
\cite{nozieres85:_crossover}, but is complicated
by the effects of a dynamic boson field and by two-dimensionality.

The total density is 
\begin{eqnarray}
  \label{eq:9}
  \rho
  &=& 
  \frac{n}{2} \left(
    1 -  \frac{\epstil}{E} \tanh(\beta E)
  \right)+ 
  \frac{\left|\psi_0\right|^2}{A} 
  \nonumber\\
  &+&\int_0^K (dk)^2 \left( 
    f(\xi_1) + f(\xi_2) - f(2E) + g(k)
  \right)
\end{eqnarray}
when condensed, with $ f(x)= -( n_B(x) + 1/2) dx /d\mu $, and $g(k)$
due to $\delta\omega$; and
\begin{equation}
  \label{eq:11}
  \rho
  =
  \frac{n}{1+e^{2\beta\epstil}}
  +
  \int_0^K(dk)^2 \left( 
    f(E_+) + f(E_-) - f(2\epstil)
  \right) 
\end{equation}
with $  f(x)=- n_B(x) dx/d\mu$ when uncondensed.

Considering the terms in eq.~(\ref{eq:9}), these can be identified as
exciton density and coherent photon density on the first line.
Under the integral are the occupation of upper and lower polaritons,
and subtraction of exciton density, required as excitons would
otherwise be counted twice.
The cutoff $K$ regularises the ultraviolet divergence.
The theory is not renormalisable, due to the constant coupling between
high energy photon modes and the excitons; but for low energy
properties, changing the cutoff introduces only small corrections.

%
However, as well as changing density, fluctuations change the critical
chemical potential by depleting the condensate\cite{popov:_chap6}.
Therefore, it is necessary to make a separate estimate of the
condensate density including fluctuations.
For a two dimensional system, this can be achieved by calculating the
superfluid response, which requires comparing the transverse (normal)
response, and the total density:
\begin{eqnarray}
  \label{eq:10}
  \rho_s&=&\rho_p-\rho_n
  , \quad
  \rho_n=\frac{1}{m \beta A} 
  \lim_{q\rightarrow 0} \left[
    \left(\delta_{ij} -\frac{q_i q_j}{q^2}\right)
    \chi_{ij}(q)
  \right]
  \nonumber
  \\
  \chi_{ij}(q)
  &=&
  \mathrm{Tr}\left( \frac{k_i k_j}{2}
    \mathcal{G}\left(\omega,k-\frac{q}{2}\right) \sigma_3 
    \mathcal{G}\left(\omega,k+\frac{q}{2}\right) \sigma_3 
  \right)
\end{eqnarray}
where $\mathcal{G}$ is the photon greens function, and $\rho_p$
is the total photon density.
The transverse response may be safely calculated without vertex
corrections, as these affect only the longitudinal response.
Combining the results of eq.~(\ref{eq:9}) and eq.~(\ref{eq:10}), the
phase transition including fluctuations occurs when $\rho_s=0$.
As the system is two dimensional, one ought to consider a
Kosterlitz-Thouless transition, at finite $\rho_s$, but this
introduces only small corrections.
The phase boundaries for four different photon masses are shown in
figure~\ref{fig:phasediag}.

%
At low densities, the phase transition occurs as the condensate is
depleted by occupation of the phase mode.
Therefore, the phase boundary has the form $T_c \propto
\rho/m_{\mathrm{pol}}$ where $m_{\mathrm{pol}}$ is the polariton mass,
which, at resonance, is just $2 m$.
This compares to the mean field boundary, which at resonance is
\begin{equation}
  \label{eq:12}
  T_c =  
  g\sqrt{n} \frac{\sqrt{1-2\rho}}{2 \tanh^{-1}(1-2\rho) } 
  \approx
  \frac{g \sqrt{n}}{-\ln \rho},
\end{equation}
the second equality being valid for small densities.  
This temperature is effectively constant on the scale of the B.E.C.
phase boundary.

As the density increases, the phase boundary including fluctuations
will approach the mean field boundary.
Due to the slow dependence of the mean field boundary on density, this
will occur at $T \approx g\sqrt{n}$, unless mass is very large.
The exact temperature where this occurs depends on the mass.
For a small mass the fluctuation boundary follows the mean field
boundary over a wider range of densities.  
The rescaled photon mass, $m^{\ast}$, provides
a small parameter near $T \approx g\sqrt{n}$, $\rho \approx n$.
Physically, at a temperature scale of $g\sqrt{n}$, the fluctuation
modes are occupied to the top of the lower polariton branch, and
occupation of the upper branch becomes relevant.

At yet higher densities, the fluctuation result again differs from
the mean field.
For $\rho > n$ the mean field theory will always predict a condensate.
However, at high densities, the system is dominated by incoherent
photons, not described in the mean field theory.
%
%
In this limit the phase boundary is that for B.E.C. of (massive) photons.

To conclude, we have calculated the spectrum of fluctuations for a
model of localised excitons interacting with propagating
photons.
In the condensed state, the fluctuations about mean field include the
Goldstone mode, and the linear dispersion causes a characteristic
change to the momentum distribution of photons.
Both the narrowing of the $N(k)$ distribution and the gap between
coherent and incoherent luminescence would provide evidence for
polariton condensation.
Fluctuations alter the form of the phase transition.
%
At low densities the transition has the $T_c\propto\rho$ form expected
for Bose condensation of pre-formed Bosons.
As the density increases, for small masses, the mean field interaction
driven limit is recovered, for densities $\rho \approx n$.
However, for yet higher densities, the boundary changes again, to Bose
condensation of weakly interacting photons.

We are grateful to B.~D.~Simons for useful discussions, and for
support from the Cambridge-MIT Institute (JK), Sidney Sussex college,
Cambridge (PRE), Gonville and Caius College Cambridge (MHS)
and the EU Network HPRN-CT-2002-00298.

\end{document}